# Dynamic Clustering in Object-Oriented Databases: An Advocacy for Simplicity

J. Darmont[1], C. Fromantin[2], S. Régnier[2+3], L. Gruenwald[3], M. Schneider[2]

| [1]E.R.I.C. | [2]L.I.M.O.S. | [3]School of CS |
|---|---|---|
| Université Lyon 2 | Université Blaise Pascal | University of Oklahoma |
| 69676 Bron Cedex, France | 63177 Aubière Cedex, France | Norman, OK 73019, US |
| `jdarmont@univ-lyon2.fr` | `michel.schneider@isima.fr` | `ggruenwald@ou.edu` |

**Abstract.** We present in this paper three dynamic clustering techniques for Object-Oriented Databases (OODBs). The first two, *Dynamic, Statistical & Tunable Clustering* (DSTC) and *StatClust*, exploit both comprehensive usage statistics and the inter-object reference graph. They are quite elaborate. However, they are also complex to implement and induce a high overhead. The third clustering technique, called *Detection & Reclustering of Objects* (DRO), is based on the same principles, but is much simpler to implement. These three clustering algorithm have been implemented in the Texas persistent object store and compared in terms of clustering efficiency (i.e., overall performance increase) and overhead using the *Object Clustering Benchmark* (OCB). The results obtained showed that DRO induced a lighter overhead while still achieving better overall performance.
**Keywords:** Object-Oriented Databases, Dynamic Object Clustering, Performance Comparison.

## 1. Introduction

Object-Oriented Database Management Systems (OODBMSs) always showed performance problems when compared to Relational Database Management Systems (RDBMSs). They really won an edge over RDBMSs only in niche markets, mainly engineering and multimedia applications. This performance problem is essentially caused by secondary storage Input/Output (I/O). Despite numerous advances in hard drive technology, I/Os still require much more time than main memory operations. Several techniques have been devised to minimize I/O transfers and improve the performances of OODBMSs, like query optimization, indexing, buffering, or clustering. Object clustering is a collaborative research topic at Blaise Pascal University (BPU) and the University of Oklahoma (OU) since the early 90's. The principle of clustering is to store related objects close to each other in order to maximize the amount of relevant information returned when a disk page is loaded into the main memory.

Early clustering methods were static [1, 2, 12, 13, 14, 15], i.e., objects were clustered only once at creation time. With these methods, modifying the placement to suit changes in data usage necessitates reorganizing the whole database. This is a heavy task that can only be performed manually when the system is idle. To support databases that are intended to be accessible on a 7 days a week / 24 hours a day basis (e.g., web-accessed databases), dynamic clustering techniques that cluster and recluster objects automatically and incrementally have been designed both by researchers and OODBMS vendors. However, since publications by the latter are very few and research proposals are not always implemented or evaluated, it is hard to select the best technique in a given context.

The objectives of this paper are to propose an overview of the research dealing with dynamic object clustering techniques; to present two methods designed at BPU and OU called DSTC and StatClust, as well as a new one called *Detection & Reclustering of Objects* (DRO); and to compare these techniques in terms of efficiency and clustering overhead. These comparisons have been performed on the Texas system using the OCB benchmark [9], which has been specially designed to evaluate clustering algorithms.

The remainder of this paper is organized as follows. Section 2 establishes a state of the art regarding dynamic clustering techniques. Section 3 presents DSTC [3], StatClust [10], and eventually details DRO. Section 4 presents the performance evaluations we performed on Texas. We finally conclude the paper and discuss future research issues.

## 2. Related Work: Dynamic Object Clustering Methods

Most dynamic object clustering methods have been motivated by needs in engineering applications like CAD, CAM, or software engineering applications. A first class of clustering strategies is based on the analysis of database usage statistics. Chang and Katz [5] proposed a physical clustering method based on a particular inheritance link called *instance to instance* and the declaration of estimated access frequencies associated with three types of relationships (aggregation, equivalence, version). The idea is allowing inheritance of data along any type of attribute and particularly along inter-object links. For instance, it is interesting, when a new version of an object is created, to automatically make it inherit from its ancestor's aggregation links toward other objects. Inherited data are stored only once, which allows an important gain in terms of disk space, but forces a physical object to be placed as close to inherited data as possible. The access frequencies and the computation of inherited attributes costs help identifying the destination page of a newly created object. If the target page is full, the system can either split the page or elect the next best page as a target page. Dynamic clustering is also coupled with an appropriate buffering strategy that is a variation of *Least Recently Used* (LRU) allowing a better usage of existing clustering. It is based on prioritizing all pages in memory. Frequently used pages have their priority increased along with their structurally related pages, while unused pages have their priority decreased with time. This method has never been implemented, except within simulation models [5, 8, 10] that hint a potential increase in performance of 200% under certain conditions.

Another method based on statistics has been proposed by McIver and King [17], who advocate that object placement determination phases must be independent of the actual placement. The strategy leans on the exploitation of three modules running concurrently. The *statistics collection module* collects general database usage statistics and also selective database usage statistics concerning depth-first or breadth-first traversals, which are assimilated to navigational and associative accesses, respectively. The *cluster analysis module* uses a variation of the Cactis algorithm [12]. It first finds out the most referenced object in the database. Then, objects linked to it are grouped on the same disk page in depth-first, by decreasing order of co-usage frequency. An advised variation is to use depth-first traversals when navigational accesses are preponderant and breadth-first traversals when associative accesses are preponderant. The type of access to select is provided by usage statistics. Clustering analysis is triggered after collection of a significant amount of statistics. The *reorganization module* rearranges objects on disk so that the database physical organization corresponds to the page assignments suggested by clustering analysis. A reorganization phase is not always necessary after each clustering analysis phase. When a reorganization phase is triggered, it deals only with objects that have not been clustered. The performance of this method has been evaluated by simulation using the *Trouble Ticket Benchmark* [16]. This study shows that the collected statistics and the proposed clustering are pertinent, and that a high overhead is caused by the database reorganization phases, where the entire database is locked and the transactions are postponed.

Cheng and Hurson state that existing strategies are generally based on one single clustering criterion [7]. Their *multi-level clustering* allows clustering objects using several criteria at once. The method associates a criterion to each of three types of relationships identified by [5]: equivalence, aggregation, and version. A proximity degree between two objects can be elaborated using the values of these criteria. Clustering is recommended when this proximity degree is sufficiently small. The clustering algorithm actually orders objects on the basis of their proximity degree. Clustering is performed by the system, without any external intervention. Furthermore, this strategy is backed up by a cost model that evaluates the benefit of a possible dynamic reorganization. This proposal has never been implemented.

Finally, an innovative strategy has been proposed to handle object clustering in the EOS distributed system [11]. This method exploits the system's garbage collector and induces a very low overhead. Clustering specifications are provided by the database administrator, who weights arcs in the class aggregation graph according to estimated access probabilities. Objects are clustered with their stronger weighted parent when created. Placements are re-evaluated afterward by the disk garbage collection process and may be modified asynchronously. This proposal has not been implemented. The authors do provide elements regarding feasibility and low cost, but this technique is intimately related to the presence of a disk garbage collector continuously working, which is costly and thus not much used in existing OODBMSs.

## 3. Studied Dynamic Clustering Algorithms

### 3.1. DSTC

DSTC is actually both a dynamic object clustering policy and its associated buffering policy, which aims at clustering together objects that are used together at near instants in time [3]. It measures object usage statistics, while respecting the following constraints: minimize the amount of data managed, maximize the pertinence of collected statistics, reduce the cost of persistent storage for these data, and minimize perturbations on running transactions. This goal is achieved by scaling collected data at different levels and using gradual filters on main memory-stored statistics. Hence, it is possible to store on disk only presumably significant statistics.

Database usage statistics concern object access frequencies and inter-object reference usage frequencies. All types of links are considered as physical references, whether they are structural links built at the schema level or logical links depending on applications or induced by physical object fragmentation. All physical accesses from one object toward another are detected and counted. Physical object reorganization is started by a trigger mechanism. Object disk storage is organized through an ordering algorithm that builds linear sequences of objects that capture "attraction forces" between objects. This sequence is sequentially transcribed in a cluster, i.e., a contiguous disk segment of variable size. The underlying algorithm was inspired by [7]. Flexibility in this approach is achieved through various parameters allowing the adaptation of system reactivity to database behavior. These parameters are set up by the database administrator. The DSTC strategy is organized into five phases.

1. *Observation phase:* During a predefined *observation period*, object usage statistics are collected and stored in an *observation matrix* in main memory.
2. *Selection phase:* Data stored in the observation matrix are sorted and filtered. Only significant statistics are retained.
3. *Consolidation phase:* Results from the selection phase are used to update data collected in previous observation phases, which are stored in a persistent *consolidated matrix.*
4. *Dynamic cluster reorganization:* Statistics from the consolidated matrix are exploited to suggest a reorganization of the physical space. Existing *clustering units* can be modified and new *clustering units* can be created.
5. *Physical database reorganization:* Clustering units are eventually used to consider a new object placement on disk. This phase is triggered when the system is idle.

The principle of the buffering management associated with DSTC is the following. When an object belonging to a cluster is accessed, the whole cluster is loaded. This avoids useless I/Os since objects in the cluster have a good probability to be used by the current transaction. A page replacement algorithm named LRU-C is also proposed. Its principle is to date clusters in the buffer rather than pages.

The DSTC strategy has been implemented in Texas [18] on Sun workstations and PCs under Linux. Performance studies have been performed with a benchmark based on OO1 [4] and baptized DSTC-CluB. They showed the efficiency of DSTC compared to a no-clustering policy on simple cases.

### 3.2. StatClust (*Statistical Clustering*)

This method extends Chang and Katz' method (see Section 2) [5]. Its authors advocate replacing user-estimated access frequencies by more reliable usage statistics [10], for each of the considered types of links (aggregation, equivalence, version). Statistics regarding read or write accesses have also been added. Clustering is automatic at object creation or update time and when a bad clustering is detected. The user can influence the clustering process through a set of parameters. A bad clustering is detected when the ratio between the number of blocks (set of contiguous pages) read in the buffer and the number of blocks read on disk is smaller than a threshold computed by the system, and the amount of collected statistics is sufficient. The detection of a bad clustering ends the collection of statistics and starts up a reclustering phase that specifies which objects might be reclustered (i.e., which objects show satisfying usage statistics). The physical placement of objects uses an algorithm close to [5], but also supports object duplication. Objects may be duplicated to increase reference locality. An object that is more read than updated is a candidate for duplication.

StatClust has been compared by simulation to static clustering techniques (ORION and Cactis) [10], but not to dynamic clustering techniques, including Chang and Katz' method, on which it is based. The results are actually very similar to those reported in [8].

### 3.3. DRO

**Overview.** The design of DRO makes use of the experience accumulated with both the DSTC and StatClust clustering methods, especially at the implementation level. Since these methods were quite sophisticated, they were also very difficult to implement properly and lots of problems occurred in the development process. Furthermore, though they attempt to minimize the amount of usage statistics stored, they use various statistical data that are not easy to manage and whose size often increases drastically. DRO is much easier to implement. It exploits both basic usage statistics and the graph of inter-object references (derived from the schema) to dynamically cluster the database. Its principle is to store together the objects that are the most frequently accessed overall. DRO has been implemented in Texas.

**Usage Statistics.** DRO stores and exploits two principal types of indicators. They are updated dynamically when the database is in use.
- The *object access frequency* measures the number of times each object is accessed. During the clustering phase, only the objects with the highest access frequencies are taken into account.

- The *page usage rate* is the ratio between the size of the data effectively stored in the page and the page size, a page being the unit of transfer between disk and memory. This ratio helps determining which pages degrade the system performance. The mean usage rate for all pages and the number of pages loaded are also computed.

The data structure presented in Fig. 1 as a UML static structure diagram is used to store DRO's usage statistics. The *PageStat* class concerns page statistics. It has three attributes: a page identifier, the number of times this page has been loaded into memory, and its page usage rate. The *ObjectStat* class concerns object statistics. It also has three attributes: an object identifier, the object access frequency, and a boolean usage indicator. The *PageObjectStat* class allows large objects to be stored on several pages. It has only one attribute: the size occupied by a given object in a given page.

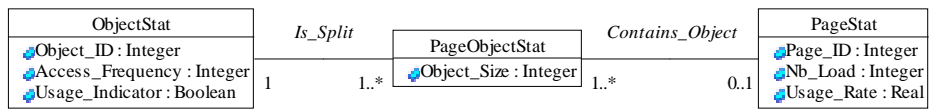

**Fig. 1.** DRO usage statistics

Whenever an object is accessed, its access frequency is incremented by 1 and its usage indicator is set to true. Page statistics are updated whenever a page moves from the main memory to disk. The statistics attached to all the objects on this page are used to compute the size occupied on the page by objects that have actually been used. The page usage rate is then computed and *Nb_Load* is increased by 1. If an object is deleted from the database, the corresponding usage statistics are also deleted. If the page that contains this object does not have any more objects in its associated *PageObjectStat* object, its statistics are also deleted. If an object is merely moved from one page to another, its usage indicator is reset to false and its link to the starting page is deleted. Its statistics will then be linked to the destination page's statistics when the object is used again.

**Clustering.** The clustering phase can be triggered manually or automatically. It is subdivided into four steps. Until physical object placement, a control procedure checks out after each step whether clustering must abort or resume.

*Step 1: Determination of Objects to Cluster.* This step helps defining the objects belonging to pages with usage rate lower than the minimum usage rate (*MinUR*) and that have been loaded in memory more times than the minimum loading threshold (*MinLT*). *MinUR* and *MinLT* are user-defined parameters. *MinUR* helps selecting pages containing a majority of unused objects or objects that are not used together. Objects stored into these pages and whose usage statistics (i.e., an *ObjectStat* object) are instantiated are selected for clustering. They are attached to instances of the *Clustering* class. Objects of class *Clustering* are linked together by two bi-directional relations called *Object_Sort* and *Object_Placement*, which store objects sorted by access frequency and a placement order of objects on disk, respectively. To proceed to step 2, two conditions must be met: a) the number of pages to cluster must be greater than one, and b) the ratio between the number of pages to cluster and the number of pages actually used is greater than the page clustering rate parameter (*PCRate*).

*Step 2: Clustering Setup.* This step helps defining a sequential placement order of objects on disk. The algorithm input is the list of objects to cluster sorted by decreasing access frequency. This step is subdivided into three phases.
- *Object clustering using inter-object references.* This first phase links objects regarding reference links. The algorithm shown in Figure 2 runs up to a user-defined maximum distance *MaxD*, i.e., the first iteration considers all the objects referenced by the starting object (distance 1), then the process reiterates for each object found, up to distance *MaxD*. When linking together objects $O_i$ and $O_j$ of access frequencies $AF_i$ and $AF_j$, the dissimilarity rate $|AF_i - AF_j| / \max(AF_i, AF_j)$ must be lower than the maximum dissimilarity rate *MaxDR* not to link objects that are too weakly bound. Objects are sorted by descending order of access frequency to generate a list defining a placement order of objects so that they can be sequentially written on disk.
- *Linking of placement order lists.* This phase links together the list parts made up in the first phase to obtain a single list. The list parts are considered in their generation order and simply concatenated.
- *Resemblance rate computation.* The third phase establishes a resemblance rate between the current object placement and the new placement proposed by the clustering algorithm. This resemblance rate helps evaluating how different the new clustering proposal is from the current physical placement of the objects. If the new cluster is found similar (for instance, if the considered objects have already been clustered), no action is undertaken. The resemblance rate is the number of objects in the proposed cluster that are not moved regarding current object placement divided by the number of objects in the cluster.

*Step 3: Physical Object Clustering.* Physical clustering is performed if the resemblance rate computed at step 2 is lower than a user-defined maximum resemblance rate (*MaxRR*). This operation clusters objects identified in the previous steps, but must also reorganize the database in order to retrieve space made available by movement or deletion of objects.

*Step 4: Statistics Update.* This update depends on a user-defined statistics update indicator (*SUInd*). If *SUInd* is set to true, all statistics are deleted. Otherwise, only statistics regarding pages containing objects that have been moved are deleted.

**DRO Parameters.** The parameters defining the behavior of the DRO strategy are set-up by the database administrator. They are recapitulated in Table 1. We obtained the default values through many experiments on Texas.

```
D = 0
End = false
While D < MaxD and not End do
   D = D + 1
   // Browse objects to cluster
   Starting_object = Clustering.Sort_first
   While Starting_object ≠ NIL and
   Starting_object.Placement_previous ≠ NIL do
      Starting_object = Starting_object.Sort_next
   End While
   While Starting_object ≠ NIL do
      Object_to_link = Starting_object
      While Object_to_link ≠ NIL and
      Object_to_link.Placement_previous ≠ NIL do
         Object_to_link = Object_to_link.Placement_next
      End while
      Found = TRUE
      While Found do
         // Find an object to cluster different from Starting_object,
         // referenced on a distance lower than MaxD, with a
         // dissimilarity rate lower than MaxDR, and attribute
         // Clustering.Placement_previous set to NIL
         Found_object = Research_procedure_result()
         If Found_object ≠ NIL then
            Object_to_link.Placement_next = Found_object
            Object_found.Placement_previous = Object_to_link
            Object_to_link = Object_found
         Else
            Found = FALSE
         End if
      End while
   While Starting_object ≠ NIL and
   Starting_object.Placement_previous ≠ NIL do
      Starting_object = Starting_object.Sort_next
   End while
End while
```

**Fig. 2.** Object clustering

| Parameter | Name | Type | Default value |
|---|---|---|---|
| Minimum usage rate | MinUR | Real | 0.8 |
| Minimum loading threshold | MinLT | Real | 1 |
| Page clustering rate | PCRate | Real | 0.05 |
| Maximum distance | MaxD | Integer | 1 |
| Maximum dissimilarity rate | MaxDR | Real | 0.05 |
| Maximum resemblance rate | MaxRR | Real | 0.9 |
| Statistics update indicator | SUInd | Boolean | True |

**Table 1.** DRO parameters

**Example of Clustering with DRO.** Let us consider the graph of inter-object references from Fig. 3 and the associated access frequencies from Table 2. With the *MaxDR* parameter set up to 0.1, Fig. 4 shows how the clustering algorithm builds an ordered sequence of objects that will be sequentially written on disk.

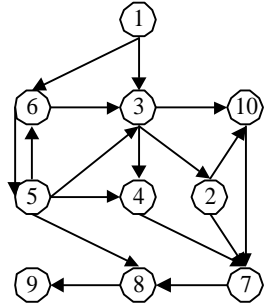

| OID | Access Frequency |
|---|---|
| 6 | 60 |
| 5 | 60 |
| 4 | 60 |
| 7 | 40 |
| 1 | 20 |
| 2 | 20 |
| 3 | 20 |
| 10 | 18 |
| 8 | 17 |

**Fig. 3.** Sample inter-object reference graph    **Table 2.** Sample access frequencies

Let *MaxD* be 1. Objects are considered by the order of access frequency. The dissimilarity rates between object couples (6, 5) and (5, 4) are both 0. The dissimilarity rates of the (6, 3), (5, 3), (5, 8), and (4, 7) couples are all greater than *MaxDR*, so the first sub-list we obtain is (6, 5, 4). The dissimilarity rate for the (7, 8) couple is 0.575 and hence greater than *MaxDR*, so (7) remains a singleton. The dissimilarity rates for the (1, 3), (3, 2), and (3, 10) couples are 0, 0, and 0.1, respectively (links to already treated objects are not considered), so the third sub-list is (1, 3, 2, 10). (8) forms the last sub-list since object #9 has never been accessed and thus must not be clustered. Now if *MaxD* is 2, we have to consider dissimilarity rates up to a "distance" (in number of objects) of 2 from the starting object. For instance, we must consider the (6, 10) couple. Its dissimilarity rate is 0.7, greater than *MaxDR*. The only change regarding the sub-lists obtained with *MaxD* set to 1 is the integration of object #8 in the (1, 3, 2, 10) sequence, because the dissimilarity rate of the (10, 8) couple is 0.0<u>5</u>, lower than *MaxDR*. Eventually, the sub-lists are merged in one list by the order of creation.

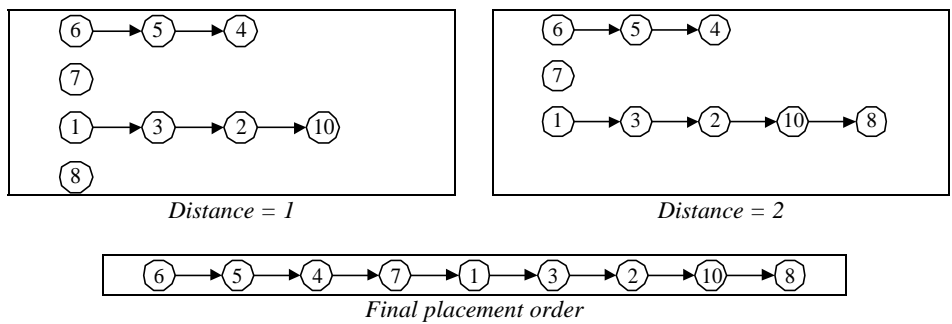

**Fig. 4.** Sample execution of the DRO clustering algorithm

## 4. Performance Comparison

### 4.1. Experiment Scope

Our initial goal was to compare the performances of StatClust, DSTC, and DRO. However, StatClust proved exceedingly difficult to implement in Texas. Since Texas exploits the operating system's virtual memory, it considers the memory buffer to be of infinite size. Thus, it is impossible to implement StatClust's module for detecting a bad clustering, because it needs to count the number of pages accessed from the disk and the buffer. Furthermore, substantial additions to Texas would be necessary to support the object replication process advocated by StatClust. Eventually, the object clustering algorithm initially builds a list of candidate pages containing objects related to the current object. To build this list, the database schema must be known. Techniques can be devised to automatically infer the schema, but none of them is easy to implement. In addition, when implementing StatClust, we found that Texas could not handle numerous transactions and the associated statistics on reasonably large databases and invariably crashed. Thus, we were not able to properly compare StatClust to the other algorithms. Hence, we only compare DSTC and DRO here.

To compare the performances of DSTC and DRO, we used a mid-sized OCB database composed of 50 classes and 100,000 objects, for a size of about 62 MB. The other OCB parameters defining the database were set to default. Two series of standard OCB transactions (1000 transactions and 10,000 transactions) were executed on this database, before and after object clustering. System performance was measured in terms of I/Os, response time, and relative performance improvement due to clustering. Only the results concerning I/Os are presented in this paper because response time plots present exactly the same tendencies and do not bring additional insight. Eventually, these experiments have been performed in several memory configurations. Since Texas makes an intensive use of virtual memory, it was interesting to see how the system behaved when the ratio *main memory size / database size* varied. The whole process was reiterated 100 times so that mean tendencies could be achieved. In each iteration, the same random seed was selected for the DSTC and DRO experiments so that they were rigorously identical.

### 4.2. Experiment Hardware and Software

The version of Texas we used is a prototype (version 0.5) running on a PC Pentium 166 with 64 MB of RAM, and version 2.0.30 of Linux. The swap partition size was 64 MB. StatClust, DSTC and DRO are integrated in Texas as a collection of new modules, and a modification of several Texas modules. Texas and the additional StatClust, DSTC and DRO modules were written in GNU C++ version 2.7.2.1.

### 4.3. Experiment Results

**DSTC.** Fig. 5 and 6 show that clustering with DSTC indeed allows a significant gain in performance, especially when the amount of main memory available is small. Clustering is definitely more useful when the database does not fit wholly within the main memory, since its effects are felt as soon as the system swaps and not only at page load time. This assumption is neatly confirmed by the clustering gain factor graph in Fig. 6. Clustering gain factor is equal to the number of I/Os necessary to execute the transactions after clustering divided by the number of I/Os necessary to execute the transactions before clustering. A discrepancy appears between Fig. 5 and 6 due to the fact that 1000 transactions are not enough: objects are used, clustered, but rarely reused following the same patterns, thus provoking an useless clustering) on small memory configurations. On the other hand, the 10,000 transaction workload appears more representative of actual database usage, allowing an average gain factor of about 2.5.

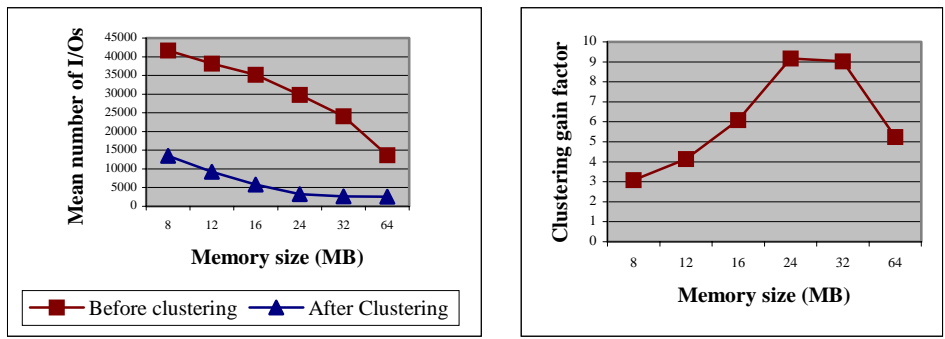

**Fig. 5.** DSTC results – 1000 transactions

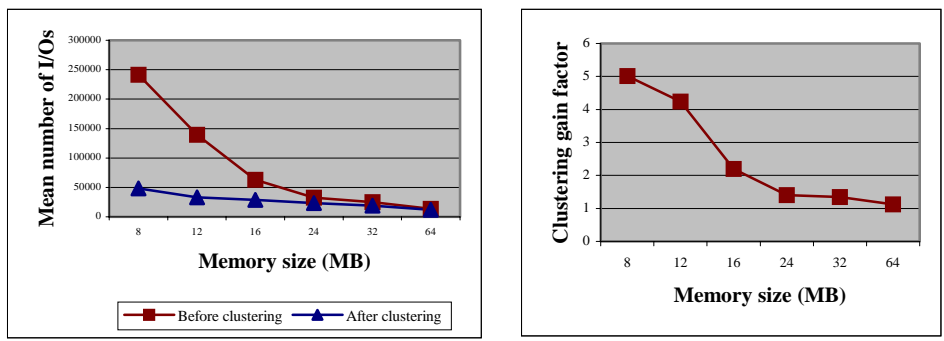

**Fig. 6.** DSTC results – 10,000 transactions

**DRO.** Fig. 7 and 8 show that DRO bears the same overall behavior as DSTC. However, the gain factor achieved with DRO on the 10,000 transaction workload looks much better. It is indeed about 15. The comparison is unfair, though, because we selected the optimal set of parameters for DRO clustering, while we could not do it for DSTC. Due to technical problems with big databases, we had to parameterize DSTC so that clustering was not the best possible. There was a threshold effect on a set of DSTC parameters. Below this "threshold", everything worked out fine but clustering was average. Beyond the "threshold", clustering units were too big for Texas to manage and the system invariably crashed.

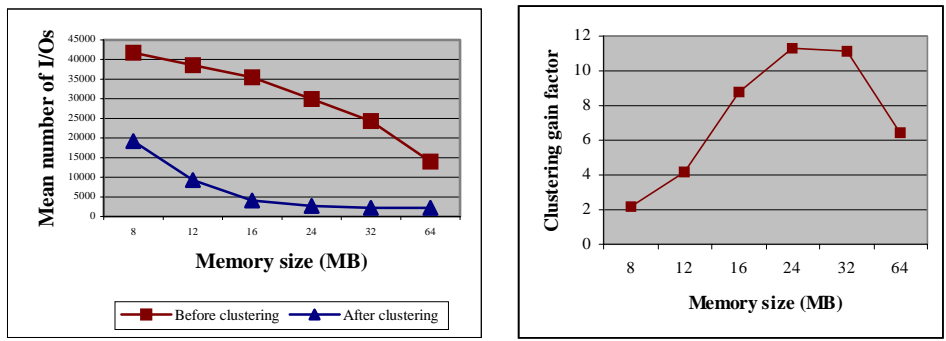

**Fig. 7.** DRO results – 1000 transactions

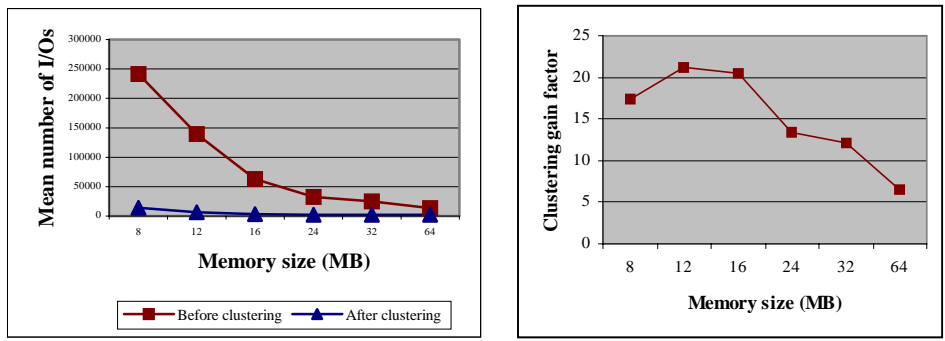

**Fig. 8.** DRO results – 10,000 transactions

**Comparison of DSTC and DRO.** To eventually compare DSTC and DRO on a fair ground, we used a smaller database so that DSTC could properly work. We used OCB's default database (50 classes, 20,000 instances, about 20 MB) and ran two series of typical transactions that were likely to benefit from clustering: depth-3 hierarchy traversals (that always follow the same type of reference) and depth-2 simple traversals (depth-first traversals). The depth of traversals was reduced regarding OCB's default parameters so that the generated clusters were not too big and the effects of clustering were clear. The traversals have been performed from 100 predefined root objects and each of them was executed 10 times.

Table 3 displays the mean number of I/Os concerning database usage before and after clustering. It shows that DSTC and DRO both achieve a substantial increase in performance (factor 6-7). DRO looks even better, though more tuning with DSTC should bring this method on the same level. Unfortunately, such tuning still provoked execution errors in Texas. The big difference between DSTC and DRO lies in clustering overhead (the number of I/Os necessary for an algorithm to cluster the database). DSTC induces a high overhead, which renders it difficult to implement truly dynamically. Its authors actually advocate its triggering when the database is idle. On the contrary, DRO, which is much simpler, present a lower overhead (about 4 times lower) and is certainly better suited to a dynamic execution.

|  | **Hierarchy traversals** | | **Simple traversals** | |
|---|---|---|---|---|
|  | *DSTC* | *DRO* | *DSTC* | *DRO* |
| Pre-clustering usage | 1682.6 | 1686 | 1682 | 1683 |
| Post-clustering usage | 270.8 | 226 | 281.75 | 236.75 |
| *Clustering gain factor* | *6.21* | *7.46* | *5.97* | *7.11* |
| Clustering overhead | 12219.4 | 3286.8 | 12174 | 2804.5 |

**Table 3.** Clustering efficiency comparison between DSTC and DRO (I/Os)

## 5. Conclusion

We have presented in this paper a representative panel of dynamic object clustering techniques, including our first effort in this field: the DSTC and StatClust techniques, which both make an intensive use of statistical counters and include clustering mechanisms with elaborated features. We have also presented a new clustering method, DRO, whose principles are based on those of DSTC and StatClust, but that is much simpler and deals with fewer statistical counters. The idea behind DRO is to provide a clustering method equivalent to or better than DSTC and StatClust while achieving simplicity of implementation.

We validated the idea that a simple dynamic clustering technique could provide better results than an elaborated one by comparing DSTC and DRO. Our results showed that DRO indeed performed better than DSTC, which could not be set up in an optimal fashion due to its inherent complexity. Furthermore, the clustering overhead induced by DRO was much lower than that induced by DSTC, definitely proving that a simple approach is more viable in dynamic context than a complex one.

To summarize, we showed that DRO was a better choice than DSTC in all circumstances. We also underlined the fact that a dynamic clustering technique is perfectly viable in an OODBMS and could achieve significant gains in performances. Since DRO is based on usage statistics, it fits well with the concept of autoadmin databases that is currently researched in major companies to automate the database tuning process [6].

The perspectives opened by this study are divided into two axes. First, the evaluation of DRO should be carried on on other systems besides Texas, which is a persistent object store rather than a full OODBMS. Such evaluations could be conducted on real OODBMSs like $O_2$, or achieved by simulation. Second, DRO itself could be improved so that clustering overhead is minimized. Some optimizations can be achieved in its code itself (at the list manipulation level, for instance), while others relate more to tuning DRO's parameters, which could also be achieved by simulation.